\documentclass[aps,prl,twocolumn,showpacs,superscriptaddress,groupedaddress]{revtex4-1}
\usepackage[T1]{fontenc}
\usepackage[latin1]{inputenc}
\usepackage[english]{babel}
\usepackage{graphicx}
\usepackage{bm}
\usepackage{amsmath}
\usepackage{amsfonts}
\usepackage{amssymb}
\usepackage{epsf}
\usepackage{epsfig}
\usepackage{subfigure}

\usepackage{draftwatermark}
\SetWatermarkAngle{45}
\SetWatermarkLightness{0.6}
\SetWatermarkFontSize{0.3cm}
\SetWatermarkScale{0.1}

\begin{document}

\title{\textbf{Optimal Control of Laser-Plasma Instabilities Using \\ Spike Trains of Uneven Duration and Delay: STUD Pulses}}

\author{B. Afeyan$^{1}$, S. H\"{u}ller$^{2}$\\
\textit {$^1$Polymath Research Inc., Pleasanton, CA, USA}\\
\textit {$^2$Centre de Physique Théorique, CNRS, Ecole Polytechnique, Palaiseau, France}\\
} \date{\today}

\begin{abstract}
Adaptive methods of laser irradiation of plasmas are proposed consisting of deterministic, `on-off' amplitude modulations in time, and intermittently changing speckle-patterns. These laser pulses consist of a series of picosecond time-scale spikes in a spike train of uneven duration and delay (STUD pulses), in contrast to hydrodynamic-time-scale modulated, multi-nanosecond pulses for laser fusion. Properly designed STUD pulses minimize backscatter and tame any absorptive parametric instability for a given set of plasma conditions, by adjusting the modulation periods, duty cycles and spatial hot-spot-distribution scrambling-rates of the spikes.
Traditional methods of beam conditioning are subsumed or surpassed by STUD pulses. In addition, STUD pulses allow an advance in the control of instabilities driven by spatially overlapped laser beams by allowing the spikes of crossing beams to be temporally staggered. When the intensity peaks of one fall within the nulls of its crossing beam, it allows an on-off switch or a dimmer for pairwise or multi-beam interactions. 

\noindent

\pacs{42.65.-k, 42.25.Bs, 42.60.-v, 52.35g, 52.35Fp, 52.35.Mw, 52.38.-r, 52.38.Bv, 52.57.-z}

\end{abstract}

\maketitle

\noindent


 In order to achieve laser fusion in the laboratory via central hot spot ignition, a great number of physical processes have to be coordinated and synchronized. One must heat a tiny fraction of thermonuclear fuel to sufficient densities and temperatures while it is being assembled by shock-wave-triggered compression, so that fusion reactions may be initiated therein, giving rise to a burn wave propagating outward through the surrounding colder fuel, heating it, and thus producing sufficient overall fusion gain. These are the goals of  inertial confinement fusion (ICF) and inertial fusion energy (IFE)~\cite{Atzeni04}. None of it is possible if the energy of the driver, in this case an array of lasers, does not correctly and symmetrically couple its energy into the target due to nonlinear optical, coherent, multi-wave interactions, known as parametric or laser-plasma instabilities (PI, LPI)~\cite{Kruer03}.  
 These efforts are compromised if plasmas irradiated by lasers back reflect substantial fractions of the laser energy thus wasting it, or deflect and spread the light around the target where it was not intended to go, thus  disrupting symmetry, or create large amplitude plasma waves which generate hot electrons and ions that preheat the fuel~\cite{Goldstein12}. Without overcoming these LPI obstacles, laser fusion will remain out of reach. Here, we propose a flexible and adaptive scheme by which LPI could be effectively tamed. It prescribes that multi-nanosecond (ns) pulses be composed of a series of picosecond (ps) time scale spikes, with gaps in between of comparable duration. The intensity contrast between on and off states ideally should be larger than  two orders of magnitude. This new scheme is the key element of  the spike trains of uneven duration and delay (STUD pulse) program announced in~\cite{Afeyan09,Afeyan12,Huller12} and introduced in this Letter. The physical mechanisms that allow controlled taming of LPI are (i) curtailed amplification or gain within a laser hot spot, (ii) damping of previously driven waves, if and when left uncoupled during successive spikes, in between hot-spot-spatial-recurrences, and (iii) scrambling of the hot spot patterns so that the probability of spatial overlap (or recurrence) of hot spots in consecutive realizations is kept low. 

In the STUD pulse program, no prior knowledge is assumed regarding (evolving) plasma conditions. We suggest relying instead on a series of successive target shots, ideally with a high repetition rate laser system, where the key elements of STUD pulse sequences are methodically varied and their relative performances measured in great detail. The goal is to discover by spectrally and ps time-scale resolved measurements of reflected light, what the optimum STUD pulse parameters must be during a particular stage of the macroevolution of the plasma conditions. These minima in reflectivity or hot electron production can be used to stitch together longer pulses and compose a whole suite that allows the navigation of the treacherous waters of ICF plasmas. There is no substitute for multiple and flexible STUD pulse interrogating, and detailed, precision diagnosing of plasmas. Here,  we shall describe salient features of STUD pulses and motivate optimal choices of parameters to urge the execution of such experiments. We will state the relevant parameters for generic PI, applicable to stimulated Brillouin and Raman scattering (SBS and SRS, which are electromagnetic scattering off  ion acoustic waves, IAW, and  electron plasma waves, EPW, respectively)~\cite{Kruer03}. 
We propose methods of STUD pulse design for primary PI mitigation and evaluate the relative performance of different STUD pulses and comment on their comparison to traditional methods of beam smoothing such as: Random Phase Plates (RPP)~\cite{Dixit93}, Smoothing by Spectral Dispersion (SSD)~\cite{Garnier97}, which is a modest improvement over RPP, and Induced Spatial Incoherence (ISI)~\cite{Lehmberg00}, which is a natural limit of STUD pulses.  We will show results from 2D fluid simulations for SBS using the code HARMONY~\cite{Huller06, Huller12}. We will follow the space-time dynamics of the interaction between incident and scattered light waves, described paraxially, and ion acoustic waves in a plasma with a flow velocity gradient, with pump depletion as the dominant nonlinear saturation mechanism.  
 
 A STUD pulse is characterized by prescribing for each spike (i) a duty cycle, $f_{\rm{dc}},$ which is the ratio of the `on' time divided by the `on+off' time of one spike, (ii) a modulation period, $\tau_{\rm{spike}}$, which is the duration of the actual  `on' time of an `on-off' pair, and (iii) a spatial scrambling rate, quantified by the number of successive spikes that elapse before the RPP pattern is replaced. It is denoted by a multiplicative factor, $\times n_{\rm{scram}}.$ 
The RPP pattern is never changed for $\times\!\infty,$ while for $\times\!1,$ for each successive spike an independent, identically distributed  RPP pattern is used.  When low frequency daughter waves are involved (such as for SBS, and for secondary instabilities following SRS, such as the Langmuir Decay Instability (LDI)), sharp rising or falling edges of identical `on' spikes can cause the creation of non-negligible amplitude harmonics which could themselves resonantly drive other low frequency waves. Since this is undesirable, a random jitter is introduced to each spike, in amplitude and in width, keeping their product fixed for easy accounting, so that these potential harmonics are smoothed out. Typically a $\pm 10\%$ jitter, suffices. The three distinguished limits of duty cycle are the overlapped beam friendly STUD2010 pulse (with copious off time gaps for crossing beams to occupy) which has a $20\%$ duty cycle with a $\pm 10\%$ jitter; the middle of the road STUD5010 pulse; and the ISI-approaching limit STUD8010. Note that in order to keep the energy in a given on-off pair the same (which is not necessary but convenient for accounting purposes), when the duty cycle is reduced, the peak amplitude must be increased correspondingly. So the peak intensity of a STUD2010 pulse is roughly 5 times higher than the corresponding RPP or SSD, while 5010 is twice as high.  

 
 By specifying the two parameters above (and the jitter), we have not yet specified the modulation period, $\tau_{\rm{spike}}.$ This is best done using length scales.There are three length scales whose relative ordering we must consider. One is the typical hot spot length in a speckle pattern, $L_{\rm{HS}}$. A good measure for this in LPI is $ 4 f^2 \lambda_0,$ where $f$ is the f-number of the final focusing optics~\cite{Garnier01, Huller10}. The width of a typical hot spot is $f \lambda_0.$ Another is the interaction length of the PI, $L_{\rm{INT}},$ which is most clearly defined within WKB ~\cite{MNR72,White10,Williams91}. This quantity,  $L_{\rm{INT}},$ depends on damping rates and on hot spot intensity. It can be shown in general to be \emph{the addition in quadrature} of the interaction length of the strong damping limit (SDL) and that in the weak damping limit (WDL), as given explicitly below.
 
The third crucial length scale is the distance traveled by the scattered light wave, at its own group velocity, during the `on' time of a spike,  $L_{\rm{spike}} = v_g \times \tau_{\rm{spike}}.$ If $L_{\rm{INT}} \ll L_{\rm{HS}},$ then scrambling hot spots, and healing preexisting growth sites, and the avoidance of long range order or re-amplification will proceed quite easily and effortlessly. But if instead, $L_{\rm{INT}} \gg L_{\rm{HS}},$ then the instability may be so easily excited, and nearby hot spots so closely coupled, that efforts to suppress it may be overwhelming. The critical case is when the interaction length is of the order of a hot spot length, and we are tasked with adjusting the spike time in order to best retard or eliminate the fastest growing PI. We call the ratio of the spike length to the hot spot length $l_{\rm{snip}},$ while the ratio of the interaction length to the hot spot length is $l_{\rm{INT}}.$ The STUD pulse modulation period is specified in terms of  the three lengths $L_{\rm{HS}}\!:\!L_{\rm{INT}}\!:\!L_{\rm{spike}}$ which, when divided through by $L_{\rm{HS}}$ becomes $1\!:\!l_{\rm{INT}}\!:\!l_{\rm{snip}}.$  One expects sufficiently improved performance of LPI  if one adopts the STUD pulse rule "cut a hot spot in half, $ L_{\rm{spike}} \sim L_{\rm{HS}} / 2,$" or $l_{\rm{snip}} \leq 1/2.$ More to the point is the rule to "cut a typical interaction length in half, $ L_{\rm{spike}} \sim L_{\rm{INT}} / 2,$" or $l_{\rm{snip}} \leq l_{\rm{INT}} / 2.$ Another STUD pulse rule is to use the largest $f/\#$ possible since then there is more maneuvering room. We recommend $f/20$, over $f/8,$ for instance, for LPI control. Another rule of STUD pulse design for primary instability control is that the damping time of the slower of the two driven waves be sufficiently short compared to the spatial recurrence time of hot spots.  This ensures that STUD pulses work to suppress instability even if the growth time is short compared to the damping time, since the scrambling of speckle patterns allows the accumulation of damping over many consecutive spikes which leads to the spatio-temporal `democratization' of gain.  Simulations of LPI with STUD pulses show that the RPP limit is a \emph{very strong fixed point}, one having a large basin boundary. Merely changing intensity patterns \emph{slowly} cannot lead to escape from the attraction of the RPP fixed point, which is precisely the case with SSD~\cite{Huller12} with slow sideway movement of hot spots~\cite{Lehmberg00}. ISI, on the other hand, is effective and well approximated by a particular suboptimal limit of STUD pulses, as shown below. 
How to overcome these restrictions is precisely the challenge faced by the STUD pulse program.  It should be noted that there are physical restrictions on modulation periods outside of LPI considerations. The lasing line of a glass laser, for example, has a lifetime of half a ps~\cite{Koechner03}. Given the scale of targets involved in ICF, typical hydrodynamic time scales over which plasmas respond to laser profile changes is 30-50 ps~\cite{Atzeni04}. Therefore, in ICF, STUD pulse may modulate a laser most easily within the window $1 \leq \tau_{\rm{spike}} \leq 20$ ps.

Here we give expressions for the Rosenbluth gain exponent and the interaction length in a linear inhomogeneity profile,
for any damping limit from $0$ to $\infty\!:$
\begin{equation} 
G_{MNR} = \frac{ 2 \pi  \gamma_0^2 }{|\kappa' V_1 V_2|} \, (1-\nu_1\nu_2 / \gamma_0^2)  > 0 \nonumber \\
\end{equation}

\begin{eqnarray}
L_{\rm{INT}}^{(\rm{WKB})} & = &  \sqrt{(L_{\rm{INT,\,WDL}})^2 + (L_{\rm{INT,\,SDL}})^2} \nonumber \\
L_{\rm{INT,\,WDL}} & = & (\,2\,/\sqrt{\pi}\,) \, \sqrt{\rm{Max}[0,G_{\rm{MNR}}] \, / \, |\kappa'|} \nonumber  \\
L_{\rm{INT,\,SDL}} & = & 2 \, \left(\, \nu_1 \,/\, |V_1| \,+\, \nu_2 \,/\, |V_2|  \,\right)  / \, |\kappa'|. \nonumber
\end{eqnarray}

\begin{figure}[!htb]
\centering
\includegraphics[width=.95\columnwidth]{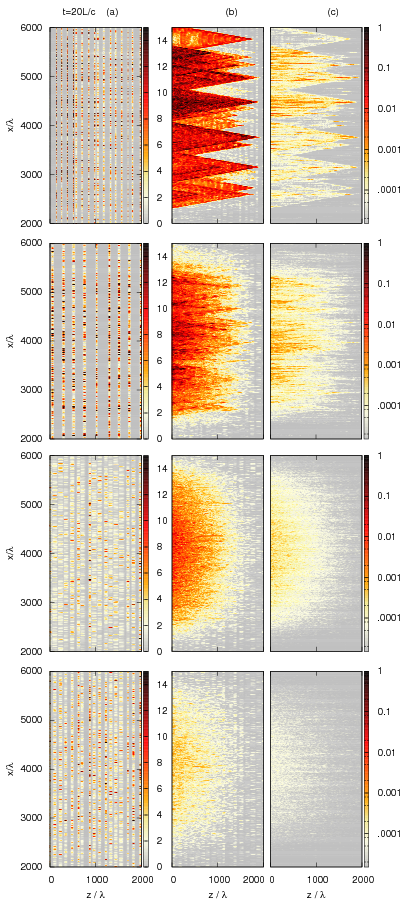}
\caption{(Color online)  Left, pump field intensity 
in linear scale, (b), center, backscattered intensity reflectivity, and (c), right, IAW density,
both in logarithmic scale,
taken at  $c \,t /L_{\rm box}\!\!=\!20.$
The combined effectiveness of varying the three primary STUD pulse parameters $f_{\rm{dc}}, l_{\rm{snip}}$ and $n_{\rm{scram}}$ is shown for 
four successively more desirable cases. In descending order, they are:
STUD$2010\!\times\!\infty,$1:1:1/4; STUD$2010\!\times\!8,$ 1:1:1/2;
STUD$8010\!\times\!1,$ 1:1:1/2;
and STUD$5010\!\times\!1,$ 1:1:1/4.
}
\label{Fig1}
\end{figure}

Here $\gamma_0$ is the primitive growth rate or the coupling coefficient of the three wave PI, $\kappa'$ is the derivative in the axial direction of the difference of the wave vectors of the three waves $\kappa' = d/dz(k_0 - k_1 - k_2).$  The group velocities and damping rates of the two daughter waves are given by $V_{1,2}$ and $\nu_{1,2},$ respectively~\cite{MNR72,White10,Williams91}. 
 
To design the simulation parameters we used the `any damping rate generic PI' formulas given above, by specializing them to SBS in the SDL~\cite{Afeyan12}:   
%
 \begin{eqnarray}
 G_{\rm{MNR}}^{(\rm{SBS})} = \frac{1.46\, (n/n_c) \, I_{14} \, \lambda_{0,\mu m}^2}{T_{\rm{e, keV}}} \left( \frac{2 \pi L_{v, 100\mu m}}{|M(0)| \, \lambda_{0.\mu m}}\right)  \nonumber \\
L_{\rm{INT}, 100\mu m}^{(\rm{SBS, SDL})} =   2 \, \frac{|1 + M(0)|}{|M(0)|}\, \left(\frac{\nu}{\omega}\right)_{\rm{IAW}} L_{v, 100\mu m} \nonumber
\end{eqnarray}
where $(n/n_c)$ is the plasma density normalized to the critical density, $I_{14}$ is the laser intensity in units of $10^{14} \rm{W/cm^2},$ $\lambda_{0, \mu m}$ is the laser wavelength in microns, $L_{v, 100\mu m}$ is the velocity gradient scale length in units of a 100 microns, $T_{\rm{e, keV}}$ is the electron temperature in keV, and $M(0)$ is the Mach number at the center of the interaction region. The damping of the IAW divided by its frequency is given by $(\nu / \omega)_{\rm{IAW}}.$ We used the value 0.1.

\begin{figure}[!htb]
\centering
\includegraphics[width=.90\columnwidth]{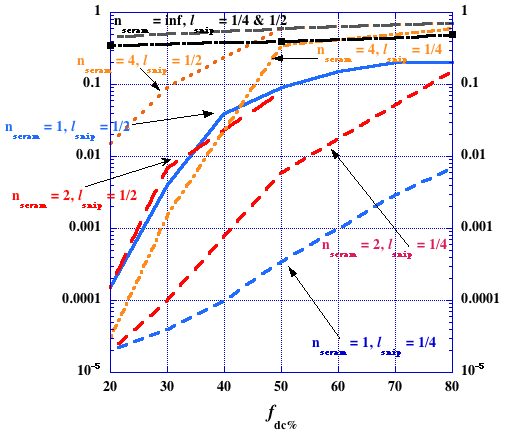}
\caption{(Color online) Design Optimization Space of STUD pulses using SBS Reflectivity (in the SDL). $R_{\rm{SBS}}$ is plotted vs duty cycle for $n_{\rm{scram}}$=1, 2, 4, $\infty,$  and for $ l_{\rm{snip}}$=.25 and .5. }
\label{Fig2}
\end{figure}


SuperGaussian spikes were used (with exponent 6) to construct a STUD pulse with a mean  contrast of 100 between on and off intensities. All simulations were done at a Rosenbluth Gain exponent of $2.14$ (in the middle of the velocity profile and at the average laser intensity). The average laser intensity was $2.67\!\times\!10^{14}\rm{W/cm}^2,$ it had an $f/8$ lens, and a Green light wavelength, $\lambda_0 = 0.527\mu m.$ The electron temperature was $2 \, \rm{keV}.$ The density was $10\%$ of critical. The flow velocity was between the Mach numbers $-3.5$ and $-2.5.$  The simulation box was $L_{\rm{box}} = 2000 \lambda_0$ long and $4096 \lambda_0$ wide. The scattered light wave was seeded at an intensity corresponding to $10^{-5}$ times the average pump intensity. This was chosen to keep pump depletion in the RPP simulations to be triggered only after an amplification with a gain exponent of $\sim11.5$. Figure 1 shows snapshots of four different cases at a time  of 70 ps ($c \, t / L_{\rm{box}} = 20$) which is 0.4 times the length of our simulations. It is at the time when asymptotic reflectivity values are reached.  
The pump intensity divided by the average intensity is the first column. The scattered light wave intensity normalized to the input plane averaged pump intensity (a reflectivity) is the second column and the density perturbation due to driven IAW w.r.t. the average density is the third column.  
The $2010\!\times\!\infty$ case approaches RPP despite its rapid temporal modulation, $l_{\rm{snip}}\!=\!1/4,$ and all the healing time needed at that small duty cycle ($f_{\rm{dc}\%}=20$). It displays repeated growth, in situ, because the speckle pattern is stationary, $n_{\rm{scram}}\!=\!\infty.$ The second row is for the same duty cycle, but $n_{\rm{scram}}\!=\!8,$ and each spike is twice as long, $l_{snip}\!=\!1/2$. It produces memory-burdened, re-amplifying, correlated-hot-spot self-organization albeit at lower levels. The third case, $8010$ is ISI-like. It does not escape memory effects despite having $n_{\rm{scram}}=1,$ and $l_{\rm{snip}}\!=\!1/2$. 
Too little `off' time is insufficient to beat memory effects. Now the STUD5010x1 case in the last row performs very well since $n_{\rm{scram}}=1, l_{\rm{snip}}\!=\!1/4,$ and it has the right duty cycle for healing and damping to occur effectively,  and to deter hot spot peaks from recurring easily. The combination of these features allows STUD pulses  \emph{to combat} the tendency to self-organize, \emph{to disrupt} percolation paths in first, scattered-light-assisted and at higher accumulated amplitudes, with IAW-assisted long-range-ordered dynamics. If $n_{scram}$ is increased to 4 or 8, these last two cases will also approach highly structured, re-amplifying, self-organized, memory-driven states, characteristic of the RPP beam fixed point. Figure 2 shows the behavior of the time-asymptotic  SBS reflectivity averaged over the transverse dimension, as a function of STUD pulse duty cycle, for the two bracketing modulation periods $l_{\rm{snip}}\!=\!1/2$ and $l_{\rm{snip}}\!=\!1/4,$ and various speckle-pattern scrambling rates, $n_{\rm{scram}} = 1, 2, 4, \infty.$ The orders of magnitude change in reflectivities observed by varying the three central parameters of STUD pulses shows the design flexibility they afford to tame LPI. Note how the widest opening in the middle of the plot allows room for compromise between degree of instability control vs. ease of experimental realizability. 
Following these STUD pulse rules, Albright et al.~\cite{Albright13}, have shown that they lead to successful SRS control even in the highly nonlinear and strictly kinetic regime of PI evolution.
 
LPI control using STUD pulses requires marshaling modern tools of Ultrafast~\cite{Weiner09}, Nonlinear~\cite{New11} and Statistical~\cite{Goodman07} optics. The requirements of optimal STUD pulses for LPI mitigation will have to be made compatible with all technical performance limits of a given laser system
~\cite{Koechner03,Haynum07}. We recommend that future high rep. rate lasers make flexible STUD pulse generation a high priority, including the staggered-peaks synchronization of crossing beams. We also recommend the Green laser option for ICF where STUD pulse mediated LPI control can unleash higher energies to be made available to drive the compression of thicker-ablator targets with higher pressures made possible with the use of higher laser intensities.  

%
We thank M. M. Fejer, A. M. Weiner, R. Salem, E. M. Campbell, M. C. Herrmann,  J. Fernandez. D. S. Montgomery, B. J. Albright, C. P. J. Barty, M. J. Edwards and  J. H. Nuckolls for insightful discussions on the potential positive  impact of the STUD pulse program and its near term implementation challenges. BA acknowledges the financial support of DOE NNSA-OFES Joint program on HEDP, and DOE OFES SBIR Phase I grants.


\begin{thebibliography}{10}
\bibitem{Atzeni04}
\newblock S.\,Atzeni and J.\,Meyer-ter-Vehn, \emph{The Physics of Inertial Fusion}, Oxford (2004).

\bibitem{Kruer03}
\newblock W.\,L.\,Kruer, \emph{The Physics of Laser Plasma Interactions}, Westview Press (2003).


%
%

\bibitem{Goldstein12}
\newblock W.\,Goldstein and R.\,Rosner, eds. Proc. Science of Fusion Ignition on NIF Workshop, LLNL-TR-570412. San Ramon, Lawrence Livermore National Security, LLC (2012). 

\bibitem{Afeyan09}
\newblock B.\,Afeyan, http://meetings.aps.org/link/BAPS.2009.DPP.\ 
TO5.7; http://www.lle.rochester.edu/media/publications/\ 
presentations/documents/SIW11/Session\_3/\
Afeyan\_SIW11.pdf

\bibitem{Afeyan12}
\newblock B.\,Afeyan, S.\,H\"{u}ller, \emph{Europ. Phys. J. Web of Conferences} (in press, 2013); also arXiv:1210.4462v1 (2012).

\bibitem{Huller12}
\newblock S.\,H\"{u}ller, B.\,Afeyan, \emph{Europ. Phys. J. Web of Conferences} (in press, 2013); also arXiv:1210.4480v1 (2012).

\bibitem{Dixit93}
\newblock S.\,N.\,Dixit, et al., \emph{Appl. Optics,} \textbf{32}, 2543 (1993).
%

\bibitem{Garnier97}
\newblock J.\,Garnier, \emph{J. Opt. Soc. Am.}, \textbf{A14},  1928 (1997).

\bibitem{Lehmberg00}
\newblock R.\, A.\, Lehmberg and J.\, Rothenberg,  \emph{J. Appl. Phys.}, \textbf{87}, 1012 (2000) and references therein.

\bibitem{Huller06}
\newblock S.\,H\"{u}ller, et al., \emph{Phys. Plasmas}, \textbf{13}. (2006).


 
 \bibitem{Garnier01}
 \newblock J.\,Garnier, \emph{Phys. Plasmas} \textbf{8}, 4914 (2001).

 \bibitem{Huller10}
 \newblock S.\,H\"{u}ller, A.\,Porzio, \emph{Laser Part. Beams} \textbf{28}, 463 (2010).
 

\bibitem{MNR72}
\newblock M.\,N.\,Rosenbluth, \emph{Phys. Rev. Lett.}, \textbf{29}, 565 (1972).

%

\bibitem{White10}
\newblock R.\,B.\,White, \emph{Asymptotic Analysis of Differential Equations}, Imperial College Press (2010).

\bibitem{Williams91} 
\newblock E.\,A.\,Williams, \emph{Phys. Fluids,} \textbf{B3}, 1504 (1991).


\bibitem{Koechner03}
\newblock W.\, Koechner and M.\,Bass, \emph{Solid-State Lasers: A Graduate Text}, Springer (2003).

\bibitem{Haynum07}
\newblock C.\,A.\,Haynum et al., \emph{Appl. Opt.}, \textbf{46}, 3276 (2007).

\bibitem{Albright13}
\newblock B.\,J.\,Albright, Y.\,Lin and B.\,Afeyan \emph{Phys. Rev. Lett.}, submitted (2013). 

\bibitem{Weiner09}
\newblock A.\, M.\, Weiner, \emph{Ultrafast Optics}, Wiley (2009).

\bibitem{New11}
\newblock G.\,New, \emph{Introduction to Nonlinear Optics}, Cambridge (2011).

\bibitem{Goodman07}
\newblock J.\,W.\,Goodman, \emph{Speckle phenomena in Optics}, Roberts and Company (2007).



%
%
%
%
%
%
%
%
%
%
%
%
%
%
%
%
%
%
%
%
%
%
%
%
%
 

%
%
%
%
%
%
%
%
%
 
\end{thebibliography}
\end{document}